# Zinc selenide single crystals codoped with active TM ions of chromium, cobalt and iron

Sergei Naydenov*, Oleksii Kapustnyk, Igor Pritula, Dmitro Sofronov

Institute for Single Crystals of NAS of Ukraine, 60 Nauky avenue, 61072, Kharkiv, Ukraine

*Correspondence e-mail: sergei.naydenov@gmail.com

**Abstract**   The search for new laser media with a transparency band of 2–5 microns is at the forefront of modern science and technology. Optical crystals based on ZnSe doped with active transition metal ions, whose emission spectra are shifted to longer wavelengths, offer significant potential. The development of laser materials with an absorption spectrum in the 1.5–2.3 micron band (for diode or fiber pumping) and an emission spectrum in the main atmospheric transparency band of 4–5 microns, which is of greatest interest for modern applications, offers unique practical opportunities. To address this problem, triple-doped (Cr, Co, and Fe):ZnSe crystals activated simultaneously with chromium, cobalt, and iron ions were grown for the first time. Comparative crystallographic, X-ray diffractometry, scanning electron microscopy, IR-spectrometry, and other physical studies of codoped (Cr, Co, Fe):ZnSe crystals grown via the vertical Bridgman method under high argon pressure were conducted. Features of their growth, morphology, and optical properties related to their crystal structure were discovered. The possibility of producing large homogeneous optical crystals with a controlled and uniform distribution of several activators within the crystal was demonstrated.

## 1. Introduction

The development of various optoelectronic and laser devices and systems that use or generate mid-infrared (MIR) radiation in the most important atmospheric transparency range from 2–5 microns requires the creation of functional laser materials that are efficient, compact, powerful enough, with a controlled and wide emission bandwidth, and resistant to adverse conditions and laser damage. In contrast to the visible range, there are few high-quality laser media for the MIR range. Oxygen atoms abnormally absorb radiation in this range. Therefore, the main role is transferred from oxygen-containing to chalcogenide laser materials. The most successful materials are binary AIIBVI crystals doped with isovalent (divalent) ions of transition metals [1-3].

A well-known medium for laser generation in the 2–3 micron range is the chromium-doped zinc selenide crystal Cr:ZnSe, which was first used for this purpose in Livermore in 1996 [4]. The luminescence band of the cobalt-doped Co:ZnSe crystal, which is usually used as a passive optical shutter [5], is in the range of 3–4 microns. To obtain coherent radiation in the range of 4–5 microns, active media based on Fe:ZnSe crystals doped with iron with a concentration of up to approximately ~$10^{19}$ cm$^{-3}$ are used [6, 7]. Currently, the most developed Cr:ZnSe laser crystal. However, practical needs, including those for special applications, including laser localization, laser targeting, laser telemetry and ranging, laser blinding, and laser cloaking, require shifting the emission band toward longer wavelengths of 4–5 microns to the best atmospheric transparency window. In addition, lasers operating in this range would be less vulnerable to conventional detection or suppression systems that use the near-infrared range. Unfortunately, Fe:ZnSe or Fe:ZnS crystals, which can theoretically solve this problem, currently have several drawbacks. The most significant of these is the extremely short lifetime of the excited energy level, no more than 290–380 ns, depending on the temperature and dopant concentration [8]. This does not achieve sufficiently high values of efficiency, especially in the mode of continuous generation, at room temperature without cooling the laser medium to the temperature of liquid nitrogen. In addition, the optical homogeneity, uniformity of activator distribution, and optical transparency of these crystals remain imperfect. Most of the working elements or experimental samples are made with cheaper polycrystals; see, for example, [9]. CVD-materials are generally inferior to single crystals in terms of both functional properties and the possibility of manufacturing deeply doped elements of arbitrary thickness for high-power lasers; see, for example, [10].

An effective way to increase the efficiency of laser generation of Fe:ZnSe active elements, which operate as part of compact lasers in the field and do not require cooling and complex and large pumping systems in the MIR range (the list of these lasers is limited), is to codope the ZnSe or ZnS matrix with several active ions at once with the transition to combined media such as coactivated crystals (Cr, Fe):ZnSe, see, for example, [11, 12] or (Co, Fe):ZnSe [13] and, possibly, triple-activated crystals (Cr, Co, Fe):ZnSe. In this case, at any temperature, chromium or cobalt ions effectively absorb external pumping radiation in the near-infrared region (usually in the 1.5–2.3 micron band, from a large list of existing semiconductors, fibers, or rare earth lasers) and transfer the excitation to iron ions that are generated in the 4–5-micron band, i.e., realizing internal transformation of the radiation spectrum. The Ferster mechanism of radiation-free (resonant) energy transfer, which is the basis of this interaction, requires a uniform and sufficiently deep level of tolerance so that the active ions are close to each other at a distance of 20–50 nm. The optimal proportions and conditions of such codoping, the physical picture and the consequences of the interaction of active ions in codoped ZnSe or ZnS matrices are still unknown and are studied mainly theoretically, e.g., [14]. For laser elements made from polycrystals, obtaining unambiguous answers to these questions is difficult because codopants can be incorporated into each single microcrystal in the same polycrystal in different ways. Moreover, codoped single crystals (Cr, Co, Fe):ZnSe, including large-sized crystals with a diameter of up to 40–50 mm (intended for the manufacture of wide-aperture elements of powerful lasers with a deep activator tolerance) and a length of up to 80–100 mm or more, are not produced anywhere in the world.

Currently, most Cr:ZnSe and Fe:ZnSe laser working elements are made of low-cost polycrystalline wafers, which are inferior to single crystals in several respects, excluding the low cost of the former. In particular, homogeneous diffusion (surface) doping of polycrystalline materials is problematic in the manufacturing of large-thickness elements and faces limitations in deep doping with a high activator concentration. The process of manufacturing high-quality optical materials based on a ZnSe (or ZnS) matrix with the addition of several activators at once becomes even more complicated. Deep doping of laser materials with a homogeneous distribution of several activators (with arbitrary concentrations) can be achieved by growing single crystals of sufficiently large size with an input load and distribution of all activators at once throughout the melt volume.

The advance of this work is the first step toward the development of a technology for the growth of structurally perfect and optically homogeneous multidoped (Cr, Co, Fe):ZnSe crystals via the vertical Bridgman method under high-pressure argon gas. A new idea is the use of cobalt as a coactivator together with chromium and iron ions and their complex tolerance to the ZnSe matrix. The expected character of the MIR absorption and emission spectra of the divalent cobalt ion is intermediate between the corresponding spectra of chromium and iron ions. In addition, the ionic radii of all these ions are very close to each other, and they are well integrated into the ZnSe matrix. Given this, codoping iron with cobalt instead of chromium (double doping) or together with chromium (triple doping) can facilitate the transfer of excitation to iron ions via a resonance mechanism, especially when deep doping with activators is needed.

## 2. Crystal growth and peculiarities of codoped ZnSe materials

Doped Fe:ZnSe, (Cr, Fe):ZnSe and (Cr, Co, Fe):ZnSe crystals were grown via the high-pressure vertical Bridgman method in high-purity graphite and/or glassy carbon crucibles under an external high-purity argon pressure of 25–30 bar. The technology for growing perfect crystals of binary compounds of chalcogenides and their solid solutions has been developed by us many times before [15-17]. The growth system was equipped with an automated control module, ensuring stable growth conditions and minimizing thermal and other physical fluctuations during ingot growth. The growth regime was defined by a temperature gradient of approximately 10–20°C/cm and a crucible translation rate of 1.0–1.5 mm/h. The impurity concentration of each dopant (Cr, Co, and Fe) in the starting materials and in the obtained crystals was on the order of several $10^{18}$ cm$^{-3}$, corresponding to approximately $10^{-3}$ wt.% in all the studied samples. The specified elements were introduced into the starting material in the form of high-purity metals. The grown ingots consisted predominantly of large single-crystalline blocks, and the yield of high-quality single crystals reached approximately 50% of the total mass of the initial charge. A series of single-crystal ingots 30–50 mm in diameter and 50–100 mm in length were obtained (Figs. 1-4). Optical samples of various geometries, including ingot wafers perpendicular to the crystal growth direction and/or slabs longitudinal to the crystal growth direction, were cut from these crystal ingots. The opposite faces of the optical samples used for IR-microscopy and IR-spectroscopy transmission measurements were laser-polished to optical quality, whereas the side surfaces were ground but left unpolished. No antireflective coatings were applied.

Fig. 1 shows a view of an ingot of a mono-doped (Fe36):ZnSe crystal grown via the vertical Bridgman method. The numbers in the chemical formula of the crystal here and below indicate the selected dopant concentration in the starting material, expressed in ppm (particle per million) units. The iron ion concentration in the resulting crystal was ~$3 \times 10^{18}$ cm$^{-3}$. For applications, crystals with dopant concentrations ranging from several $10^{18}$ cm$^{-3}$ to $10^{19}$ cm$^{-3}$ are most often used. At even higher activator concentrations, strong concentration quenching of luminescence can occur in crystals.

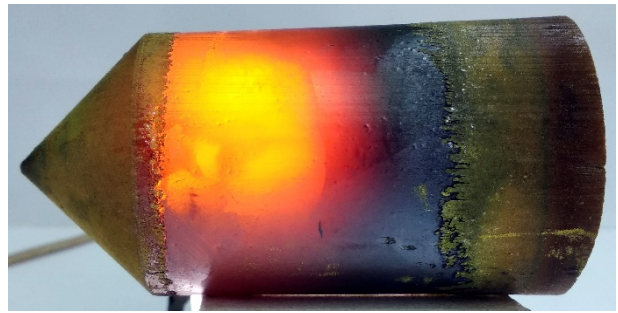

**Figure 1** The ingot of a mono-doped (Fe36):ZnSe crystal was grown via the Bridgman method.

Fig. 2 shows another ingot of doped (Fe22):ZnSe crystal grown via the vertical Bridgman method and cleaved in half along the cleavage plane. The iron ion concentration in this crystal was ~$2 \times 10^{18}$ cm$^{-3}$. It was a single-crystal ingot, on the surface of which the boundaries of several large "twins" emerged and were clearly visible. The crystal was cleaved along the boundary of one of these "twins". X-ray examination of the cleavage plane revealed that it was oriented normally [110]. It is well known that the ZnSe crystal has a sphalerite (zinc blende) structure, and this plane is the cleavage plane, i.e., the plane of the most likely cleavage of the crystal. This is caused by two factors. First, the family of parallel planes (110), together with another typical family (100), belongs to the type of the most equidistant parallel planes of symmetry in a cubic lattice. The distance between the nearest planes of an arbitrary family with integer Miller indices $(hkl)$ in a cubic lattice with a period $a$ is determined by the formula

$$d_{hkl} = \frac{a}{\sqrt{h^2 + k^2 + l^2}} \qquad (1)$$

For the (100) plane family inter-distance is equal to $d = a$, and for the "diagonal" (110) plane family it is equal to $d = a/\sqrt{2}$, i.e. approximately 1.4 times smaller. However, any (110) plane contains the same number of cations and anions (per unit area) and is generally electrically neutral, unlike the (100) planes, which are alternately charged either positively when they contain only cations (zinc ions) or negatively when they contain only anions (selenium ions). Therefore, the Coulomb interaction between the (110) family planes is the weakest in the crystal lattice. This leads to the (110) planes being cleavage planes in ZnSe crystals. Isovalent codoping does not change the charge state of the cation sublattice. Consequently, it does not change the cleavage pattern of these crystals. This is confirmed by our experiments.

It is interesting to discuss the following experimental fact. Fig. 2 clearly shows that the (110) cleavage plane of the crystal runs strictly along its bulk diagonal, i.e., at an angle of 45 degrees relative to the crystal growth axis. Consequently, when a single crystal is grown, its growth occurs primarily along the axial growth direction orthogonal to the (100) plane or along the radial growth direction orthogonal to the equivalent planes (010)~(100) when cubic symmetry operations are applied to both planes, which form an angle of 45 degrees with the (110) plane. The angle between two arbitrary planes with Miller indices $(h_1 k_1 l_1)$ and $(h_2 k_2 l_2)$ in a cubic lattice is determined by the formula

$$\cos\theta = \frac{h_1 h_2 + k_1 k_2 + l_1 l_2}{\sqrt{h_1^2 + k_1^2 + l_1^2}\sqrt{h_2^2 + k_2^2 + l_2^2}} \qquad (2)$$

From here, it is easy to prove geometrically that no other planes that form an angle of 45 degrees with the (110) plane exist in the cubic lattice. From a crystallographic standpoint, the (100) and (010) planes are the "extreme" planes with the maximum reticular density of atoms in a simple cubic lattice. The reticular density of atoms on a plane $(hkl)$ is inversely proportional to the factor $\sqrt{h^2 + k^2 + l^2}$ and decreases rapidly for planes with large Miller indices. The sphalerite structure is formed by two "nested" cubic sublattices, which are shifted along the main diagonal of the cube by a quarter of its length. Here, the bulk "diagonal" plane (111) has the maximum reticular density, which, in a simple Bravais lattice, would formally have a lower reticular density than the "diagonal" (100) or (010) plane. For sphalerite, each (100) or (010) plane contains atoms of only one specific type (they correspond to cations or anions), whereas the (111) plane contains atoms of both types. Therefore, for each of the cationic and anionic sublattices of the total crystal lattice of the binary compound ZnSe, the crystallographic planes (100) and (010) are "quasiextreme" planes with the highest reticular density of atoms of the selected type.

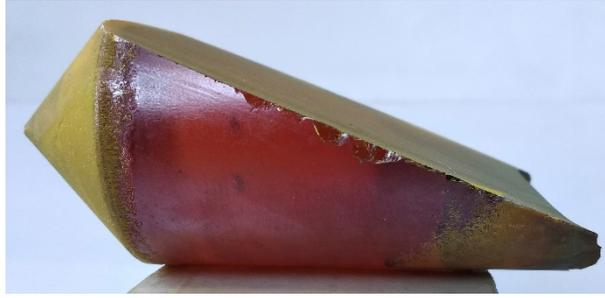

**Figure 2** The as-grown single-crystal (Fe22):ZnSe cut along the (110) crystallographic plane.

Moreover, planes with the lowest reticular density typically exhibit the highest growth rate in crystals. As a result, during the initial stage of crystal nucleus growth, these planes are overgrown much faster than those parts of the crystal that correspond to planes with higher reticular density. Crystal growth along these "fast" growth directions (corresponding to planes with high Miller indices) ceases. The bulk crystal subsequently grows primarily along the planes with the highest reticular density. Thus, in our work, we clearly observe that this physical peculiarity takes place for doped ZnSe crystals.

Unfortunately, growing large ZnSe single crystals presents significant challenges. In more than 90% of the cases, when these crystals are grown via the Bridgman melt method, the resulting ingot consists of several fused single-crystal blocks. Each of these blocks corresponds to a different nucleus, which forms randomly at the beginning of crystallization and typically at the periphery of the crystal (near the crucible walls). Suppressing excessive nucleation is crucial for growing high-quality crystals. This is achieved via various technological approaches. We achieve this by selecting the optimal growth temperature regime with a low but sufficient temperature gradient at the crystallization front and a low crystal pullout rate.

Fig. 3 shows a crystal ingot (Cr63, Fe75):ZnSe grown with double codoping with chromium and iron. Optical methods revealed that the average chromium ion concentration in the crystal is ~$1.22 \times 10^{18}$ cm$^{-3}$, and the iron ion concentration is ~$3.35 \times 10^{18}$ cm$^{-3}$. This concentration was several times lower than the dopant concentration in the starting material before crystal growth. The decrease in chromium concentration in the crystal may be due to its segregation coefficient in the melt being significantly lower than unity. As a result, chromium is displaced at the crystallization front by the growing crystal, where the chromium concentration is lower than that in the melt. This feature is observed during the growth of mono-doped Cr:ZnSe crystals. With respect to iron, during the growth of mono-doped Fe:ZnSe crystals, its concentration in the starting material and in the grown crystal is approximately the same. The sharp decrease in the expected chromium and iron concentrations in codoped crystals requires further study. This feature may be associated with the agglomeration of iron and chromium microparticles in the high-temperature melt, owing to which their coefficient of entry into the crystal in the form of isovalent substitution ions decreases sharply.

Fig. 4 shows the triple-doped (Cr125, Co100, Fe100):ZnSe crystal ingot with chromium, cobalt and iron. Optical methods revealed that the average chromium concentration in the crystal is ~3.54×10$^{18}$ cm$^{-3}$, the cobalt concentration is ~7.35×10$^{18}$ cm$^{-3}$, and the iron concentration is ~4.49×10$^{18}$ cm$^{-3}$. As in the previous dual-doping case, the concentration of chromium and iron ions in the crystal was much lower than expected. Moreover, almost all of the cobalt from the starting charge was incorporated into the matrix of the grown crystal. This feature may prove important when choosing the optimal activator concentration in the development and production of new codoped laser crystals on the basis of a zinc selenide matrix.

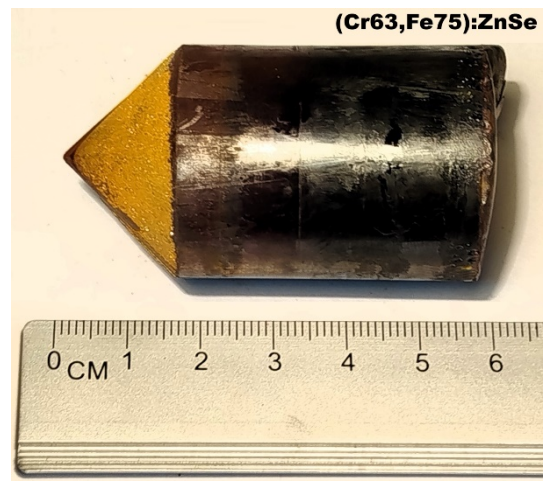

**Figure 3** The ingot of the dual-doped (Cr63, Fe75):ZnSe crystal.

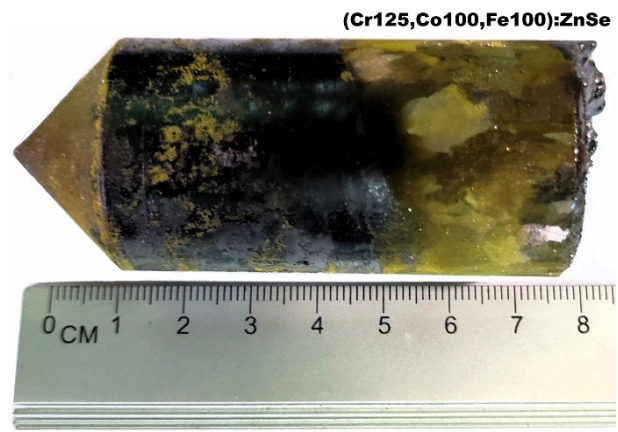

**Figure 4** The ingot of the triple-doped (Cr125, Co100, Fe100):ZnSe crystal.

When codoped laser crystals are grown, the goal is to obtain a material with the highest possible transparency and optical homogeneity, as well as a uniform distribution of activators throughout the working volume of the crystal. Unfortunately, the crystalline quality of multidoped crystals is still inferior to that of monocoped crystals. Codoped crystals contain more inclusions (gas pores, traces of external contaminants, etc.) and various structural macroscopic inhomogeneities. The optical transparency of these materials in the 5–15 micron wavelength band still falls short of the theoretical value of 70% for ZnSe (without antireflective coatings). In our opinion, this is due to the poor choice of growth conditions, which will be optimized further. However, the distribution of several coactivators within such crystals is quite uniform. This is confirmed by analysis of the IR transmission spectra (see the next section) obtained by irradiating different parts of the crystal.

## 3. Crystal structure and IR-absorption spectra

All the codoped (Cr, Co, Fe):ZnSe crystals have a sphalerite structure. We previously grew doped Cr:ZnSe, Fe:ZnSe, and dual-doped (Cr, Fe):ZnSe crystals. Triple-doped (Cr, Co, Fe):ZnSe crystals were grown for the first time. To clearly confirm their single-phase nature, we performed X-ray powder diffraction (XRD) analysis of their crystal structure and energy-dispersive spectroscopy (EDS) analysis of their atomic composition by scanning electron microscopy (SEM).

Fig. 5 shows a characteristic X-ray reflection for a sample of crystalline material taken from the central part of the (Cr125, Co100, Fe100):ZnSe crystal with high transparency in the 5–15 micron band. Powders for sample preparation were obtained after crystal cleavage and mechanical grinding of the fragments in an agate mortar under high-purity conditions. An X-ray diffractometer (Aeris Research, Malvern Panalytical B.V., Netherlands) was used to analyze the samples. An X-ray tube with a copper anode was used as the radiation source (Kα, ~1.54 Å). X-ray phase analysis results were processed via the Rietveld method [18, 19] via the HighScore (Plus) v5.0 software package [20]. The analysis results confirm the 100% single-phase nature of the obtained material with a structure matching the structure of crystalline ZnSe (complete match for 6 selected lines in the X-ray diffractometry spectrum). The calculated value of the cubic lattice period $a_{exp} = 5.6688 \text{Å}$ coincides with a high accuracy with the tabulated value $a_{tab} = 5.6676 \text{Å}$. The overestimated experimental lattice period may be due to the presence of structural defects in the crystal (primarily dislocations) and/or slight lattice sparsification caused by the substitution of larger chromium (ionic radius $r_{Cr2+} = 0.80 \text{Å}$), cobalt ($r_{Co2+} = 0.75 \text{Å}$), and iron ($r_{Fe2+} = 0.78 \text{Å}$) ions for divalent zinc ions ($r_{Zn2+} = 0.74 \text{Å}$). The atomic parameters of divalent zinc and cobalt ions are closest to each other. This may explain our experimental discovery that, upon triple doping of grown crystals, cobalt ions are the best incorporated into the zinc selenide matrix.

The positions of the peaks in Fig. 5 are well described by the theoretical dependence

$$\sin \theta_{hkl} = \frac{\lambda}{2a_{exp}} \sqrt{h^2 + k^2 + l^2} \tag{3}$$

where $\lambda$ is the wavelength of incident X-ray radiation. The intensity of the peaks is proportional to the concentration of microcrystals present in the ground powder. The greatest intensity occurs for X-rays reflected from planes with low Miller indices, such as (111), (022), (131), and symmetrically equivalent planes, which are obtained from the original planes by various cyclic permutations and changes in the signs of their Miller indices. The planes in these families are spaced from each other by a distance significantly greater than those in other families (with higher Miller indices), and these planes themselves have a higher reticular density.

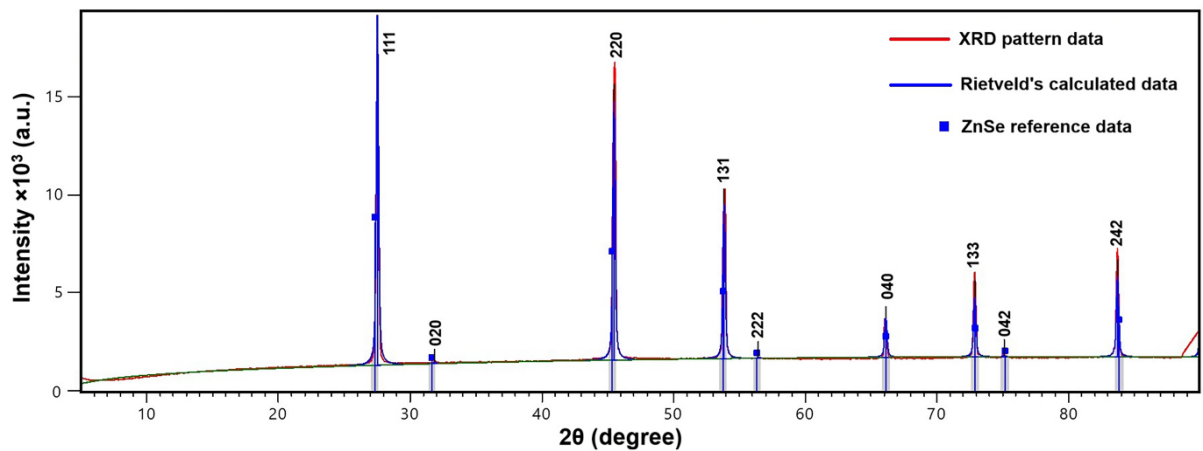

**Figure 5** X-ray diffractometry results for the (Cr125, Co100, Fe100):ZnSe crystal material. The Miller indices of the corresponding crystallographic faces are indicated for the observed peaks.

The X-ray diffraction pattern in Fig. 5 contains peaks of maximum intensity for the lines corresponding to the (111) and (022) crystallographic planes, as well as the (202) or (220) planes, which are equivalent to the (220) plane in the framework of cubic symmetry. It lacks lines corresponding to the (110), (101) or (011) cleavage planes of the ZnSe single crystal. This can be explained as follows. During the mechanical grinding of crystal fragments into powder, the crystallites are initially most likely cleaved along the (110), (101) or (011) cleavage planes in the cubic lattice. As a result, small crystallites acquire the characteristic habitus of a zinc blende crystal. However, their further grinding becomes energetically favorable when cleaving along other planes with a minimal energy of the Coulomb interaction between them. These include the family of bulk "diagonal" planes (111), which, as noted earlier, are distinguished by the largest possible distance between adjacent parallel planes in a cubic lattice (equal to the lattice period). There is also a family of planes (022), parallel to the planes of the (011) family and closest to them among all families with the maximum possible distance between the planes of the family. As directly follows from Eq. (1), the distance between the planes of the (022) family is half as much as the distance between the planes of the (011) family. As a result of the crushing of the crystallites to the finely dispersed powder stage, the fraction of the remaining microcrystals with crystallographic faces of (111) and (022) may be dominant.

The composition of the codoped (Cr, Co, and Fe):ZnSe crystals was studied via a JSM-6390LV scanning electron microscope (JEOL, Japan) with an X-Max 50 energy dispersive spectrometer (OXFORD Instruments Analytical, Great Britain). Samples in the form of longitudinal wafers ~5 mm thick were cut from the ingot and processed for SEM and EDS analysis (see Figs. 6-7). Sections corresponding to the nose, center, and tail of the ingot were selected for analysis of the wafers. Axial (along the crystal growth direction) and radial variations in the composition of the crystalline ingots were studied, as were composition variations near visible structural and optical macrodefects, inclusions, etc. High stoichiometry (within the accuracy of the EDS analysis method of ~0.5 at.%) was confirmed for all the obtained crystals, regardless of the choice of scanning location. The calculated ratio of the number of atoms [Zn]/[Se]~0.99 is maintained in almost all crystal parts, even near captured gas pores and/or inclusions of external impurities (presumably, carbon contamination), where the absolute concentration of zinc and selenium decreases several times and becomes comparable with the concentration of external impurities.

Figs. 6-7 show experimental samples cut from grown codoped crystals. The polished samples illuminated with natural light clearly reveal the large-block structure of the grown crystalline ingots, which consists of several single-crystalline blocks. Characteristic macrodefects of ZnSe-type crystals are visible, including parallel series of alternating "twins", clusters of fine-grain boundaries near the boundaries of large mono-blocks, and others. Optical images of the samples illuminated with a powerful light source (a broadband halogen lamp) reveal optical inhomogeneities and various structural defects in the crystals, which are clearly visible in the IR range. Using IR microscopy, we detected clusters of gas pores and micron-sized inclusions (growth traps) of external impurities, especially in the central regions of the codoped crystals. This central part usually remains underheated during crystal growth compared with the fields near the crucible walls because of the poor thermophysical properties of the melt. Therefore, more intensive capture of gas pores and inclusions can occur. EDS analysis of the experimental sample revealed the presence of carbon (up to several tens of atomic percent) on the pore walls and inclusions. We hypothesize that these growth traps are caused either by contamination of the starting materials with organic matter or by the transfer of graphite contaminants from the growth system components in a high-pressure argon atmosphere. Future plans include eliminating these factors and growing optically perfect crystals free of inclusions and gas pores.

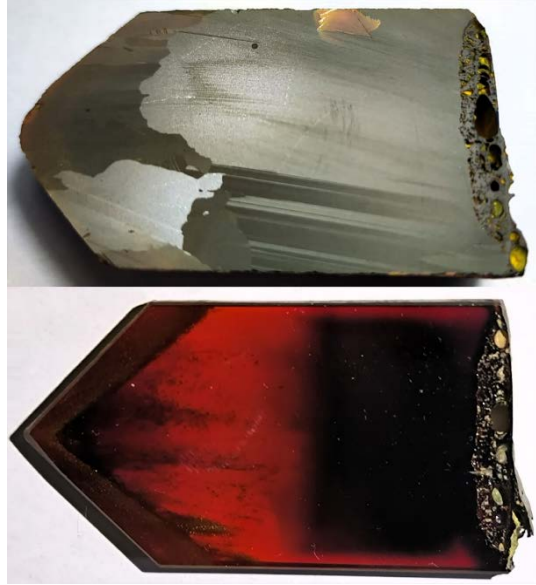

**Figure 6** The experimental sample (longitudinal wafer) was cut from the (Cr63, Fe75):ZnSe crystal without illumination (top) and with strong illumination (bottom).

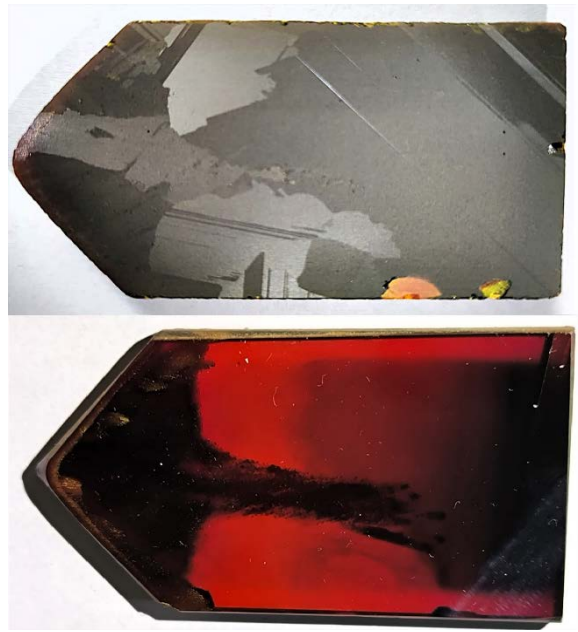

**Figure 7** The experimental sample was cut from a (Cr125, Co100, Fe100):ZnSe crystal without illumination (on the top) and with strong illumination (on the bottom).

Fig. 8 presents IR spectroscopy data for optical samples cut in the form of a longitudinal plate from the paraxial part of triple-doped (Cr, Co, Fe):ZnSe crystals. The IR-transmission spectra were obtained in the 1.3–22.2 micron band via the FTIR Spectrophotometer Spectrum One (Perkin Elmer, USA). The thickness of the prepolished crystal plate was 4.83 mm. No antireflection coating was applied. Therefore, taking into account the calculated value of the refractive index of the zinc selenide material, the theoretical transmittance limit was approximately 70% over the entire passband. The measurements were carried out in different parts of the plate, which correspond to the nose, middle, and tail parts of the crystal. All transmission spectra have three characteristic absorption maxima at wavelengths $\lambda_{Co} \approx 1.60\,\mu m$, $\lambda_{Cr} \approx 1.77\,\mu m$, and $\lambda_{Fe} \approx 3.05\,\mu m$. These peaks correspond to the IR-absorption maxima of divalent cobalt, chromium, and iron ions in a zinc selenide matrix. The first two maxima are located in the absorption bands of chromium and cobalt ions. They are very close to each other and, at low resolution, as in Fig. 8, are virtually indistinguishable. As a result, their combined transmission spectrum appears as a spectrum with a single common maximum $\lambda_{Cr-Co} \approx 1.71\,\mu m$ shifted toward the short-wavelength region relative to the chromium ion absorption maximum (or, conversely, toward the long-wavelength region relative to the cobalt ion absorption maximum). The magnitude of this shift depends on the relative concentration of chromium and cobalt ions in the crystal. Compared with that of mono-doped Fe:ZnSe crystals, the iron absorption band in triple-doped crystals remains virtually unchanged (the absorption maximum does not shift).

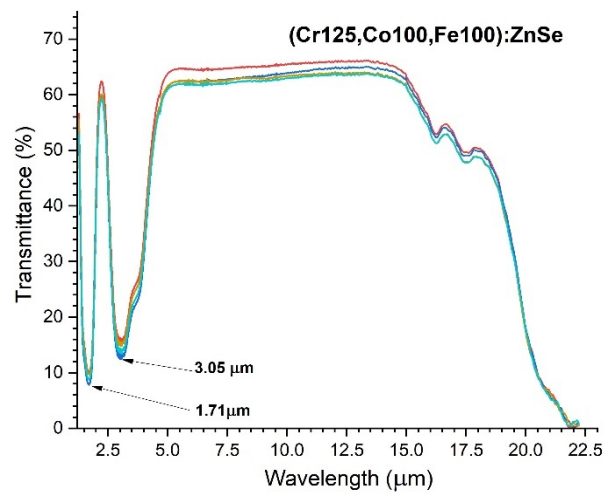

**Figure 8** Mid-IR transmission spectra of the (Cr125, Co100, Fe100):ZnSe crystal. Lines of different colors correspond to transmission spectra obtained for different parts of the crystal.

Importantly, in different regions with good optical transparency, the transmission curves are very close to each other, both at the absorption maxima and in the characteristic wide transparency band of 5–15 microns. The optical transparency of the crystal reaches 65%, which is quite close to the theoretical limit of 70%. The decrease in the IR transparency is due to the presence of various structural defects and optical inhomogeneities in the crystal, which primarily lead to light scattering. This means that the concentration of each coactivator varies slightly within the crystal, meaning that the distributions of all three activators are virtually uniform. This is a distinctive feature of codoped crystals grown via the Bridgman melt method. This enables the growth of crystals with a controlled and specified distribution of dopants within the crystal.

## 4. Conclusions

Optical crystals (Cr, Co, and Fe):ZnSe codoped with various transition metal ions were grown via the vertical Bridgman method under high argon pressure. Crystals triple-doped with chromium, cobalt, and iron were grown for the first time. The crystal structure of the obtained materials was studied via X-ray diffractometry, scanning electron microscopy, IR-spectroscopy, and other methods. The single-phase nature, stoichiometry, optical transparency, and spatial homogeneity of the activator distribution for the grown crystals were confirmed. Several experimental features of the growth and optical properties of these crystals, related to their crystal structure, were discovered and explained. The sphalerite crystal structure dictates the choice of (110) cleavage planes for these single crystals, as well as the dominant direction of single-crystal grain growth, which is primarily perpendicular to the (100) crystallographic planes and their equivalent planes in the cubic symmetry lattice. Isovalent substitution ions are readily incorporated into the ZnSe matrix, particularly cobalt ions, whose ionic radius is closest to that of zinc ions. However, the concentration of chromium and iron ions with simultaneous codoping is significantly lower than expected (several times lower than that in the growth starting material) than that with individual monodoping of ZnSe crystals. The optical absorption spectrum of active iron ions in the 2–4 micron band remains virtually unchanged with triple doping. While the total absorption spectrum of chromium and cobalt ions in the 1.5–2.3 micron band becomes bimodal, in contrast to mono-doping with these elements, its maximum shifts by almost 100 nm to the shortwave region relative to the maximum absorption of chromium ions (or to the longwave region relative to the maximum absorption of cobalt ions). Research has stimulated the development of new laser materials based on zinc selenide single crystals with controlled and uniform multidoping with transition metals.

**Acknowledgements**   This work was supported by the National Research Foundation of Ukraine (NRFU), grant number 2025.06/0006. The authors are also grateful to Dr. Nazar Kovalenko and Mr. Igor Terzin for assistance in the experimental work and Dr. Oleg Lukienko for performing the X-ray diffractometry measurements.